\documentclass[11pt,a4paper]{article} \usepackage{jcappub}
\usepackage{graphics}  
\usepackage{epsfig}
 \usepackage{amssymb}   
 \usepackage{verbatim}

\IfFileExists{srcltx.sty}{\usepackage[active]{srcltx}}

\newcommand{\be}{\begin{equation}}
\newcommand{\ee}{\end{equation}}
\newcommand{\bea}{\begin{eqnarray}}
\newcommand{\eea}{\end{eqnarray}}

\begin{document}

\title{Extreme gravitational waves from inflaton fragmentation}

\author[a,b]{Anupam Mazumdar}

\affiliation[a]{Physics Department, Lancaster University, Lancaster, LA1 4YB, UK}

\affiliation[b] {Niels Bohr Institute, Blegdamsvej-17, Copenhagen, DK-2100, Denmark }

\author[c]{Ian M. Shoemaker}
\affiliation[c]{Theoretical Division, MS B285, Los Alamos National Laboratory, Los Alamos, New Mexico 87545, USA }

\emailAdd{a.mazumdar@lancaster.ac.uk}
\emailAdd{ianshoe@lanl.gov}


\abstract{
Although inflationary models generically predict a flat spectrum of gravitational waves, we point out a general process that produces a sharply peaked spectrum of gravitational radiation.  This process is generic for inflationary models with a complex inflaton field which couples to fermions.  In particular, for chaotic models these may be the most extreme gravitational waves in the Universe with a very large abundance $\Omega_{GW} h^2\sim 10^{-9}$ and ultra-high frequency, $10^{10}$~Hz.  Although not amenable to space based interferometers, the signal from this model may be detectable by future table top experiments. 


}

\arxivnumber{1010.1546}

\maketitle
\section{Introduction}

In addition to providing solutions to the horizon and flatness problems, inflation is also responsible for the generation of density perturbations and reheating the universe.  The excitation of both scalar and tensor perturbations on very large scales with a nearly flat spectrum is a generic consequence of inflation. For the 
highest scale model of inflation, $\rho\sim 10^{64}~({\rm GeV})^4$, it is possible to obtain a large tensor to scalar ratio, $r\sim 0.1-0.001$, which can potentially be observable with the PLANCK satellite~\cite{Mazumdar:2010sa}.

Lack of any positive detection of stochastic gravitational waves (GWs) suggests that the scale of inflation is lower than the upper limit $10^{64}~({\rm GeV})^4$. Chaotic inflation with super-Planckian VEVs is perhaps one of the best known examples of high scale which is compatible with the current CMB data~\cite{Hinshaw:2008kr}. In this scenario, one has a simple potential given by $V= m^2\phi^2$ where $m\approx 10^{13}$~GeV is the mass of the inflaton field $\phi$~\cite{Linde:1981mu,Linde:1983gd}.  Inflation occurs at super-Planckian VEVs, $\phi\approx 5-10 M_{P}$ (where $M_{P}= 2.4\times 10^{18}$~GeV), and ends near $\phi\approx M_{P}$, and can potentially lead to long wavelength stochastic gravitational waves~\cite{Mazumdar:2010sa}.

However, there is another interesting source of generating GWs connected intimately with the end of inflation. The end of inflation gives rise to an epoch of reheating which can be responsible for generating GW, albeit at short sub-Hubble scales, with a  spectrum which would be peaked as compared to a flat spectrum. The process of reheating is essential for generating thermal entropy and the creation of the observed degrees of freedom. After inflation, it is typically assumed that the inflaton oscillates coherently and eventually decays through ordinary perturbative processes ($2$ or $3$ body decays of the inflaton)  or  non-perturbatively (for a review see~\cite{Allahverdi:2010xz}). However a novel possibility has been suggested in which inflaton decay could be delayed if the inflaton forms non-topological solitons of its own matter: inflatonic $Q$-balls~\cite{Enqvist:2002si,Enqvist:2002rj}, which subsequently decay via surface evaporation~\cite{Coleman:1985ki}. Such fragmentation leads generically to a very inhomogeneous and anisotropic distribution of matter.  If the matter  undergoing fragmentation once dominated the energy density of the Universe, then the present-day abundance can be very large.  This idea has been explored in the context of supersymmetric flat direction condensates originally in~\cite{Kusenko:2008zm,Kusenko:2009cv} and more recently by~\cite{Chiba:2009zu,Chiba:2010ff}.~\footnote{The Q-balls are formed from the fragmentation of the inflaton condensate. The fragmentation of any scalar condensate was first described in the context of gauge mediated supersymmetry breaking~\cite{Kusenko:1997si,Kusenko:1997zq} and subsequently in gravity-mediated models~\cite{Enqvist:1997si,Enqvist:1998ds}. In these supersymmetric models the scalars are made up of combinations of squarks and sleptons. The potential obtains quantum corrections by virtue of gauge/gaugino and fermion/sfermion couplings such that the minimum of the potential is flattened and becomes shallower than quadratic, for a review see~\cite{Dine:2003ax,Enqvist:2003gh}. The scalar condensate oscillating in such a potential gives rise to a negative pressure on average~\cite{Enqvist:1997si}, which triggers the instability in the condensate and thus the fragmentation.  The process of fragmentation has been studied analytically and numerically~\cite{Kasuya:1999wu,Kasuya:2000wx, Kasuya:2000sc,Kasuya:2001hg}. } The cosmological implications of supersymmetric $Q$-balls has been extensively studied~\cite{Frieman:1988ut,Frieman:1989bx,Griest:1989bq,Griest:1989cb,Enqvist:1997si,
Enqvist:1998ds,Enqvist:1998xd,Kusenko:1997ad,Kusenko:1997hj,Kusenko:1997it,Kusenko:1997vp,Laine:1998rg,Enqvist:1998en,
Enqvist:1998pf,Axenides:1999hs,Banerjee:2000mb,Battye:2000qj,Allahverdi:2002vy,Enqvist:2003gh,Dine:2003ax,Kusenko:2004yw,
Kusenko:2005du,Berkooz:2005rn, Berkooz:2005sf,Kusenko:2008zm,
Johnson:2008se,Kasuya:2008xp,Sakai:2007ft,Campanelli:2007um,Kasuya:2000wx,
Kawasaki:2005xc,Kasuya:2007cy,Shoemaker:2008gs,Campanelli:2009su,Shoemaker:2009kg,Kusenko:2009iz}.  It is this fragmentation process, leading to inhomogenous and anisotropic motion of matter in the scalar condensate that sources the GWs.

In this paper we point out that the fragmentation of an inflaton condensate would naturally lead to a very large amplitude GW with a high peak frequency of order $\sim 10^{10}$~Hz with a narrow spectrum which falls as $k^{-1}$ on sub-Hubble scales.  Such GWs may be detectable in foreseeable future table-top experiments~\cite{Cruise:2000za,Cruise:2006zt,Nishizawa:2007tn,Nishizawa:2008se,Akutsu:2008qv}.  Indeed the authors of~\cite{Akutsu:2008qv} have obtained an impressive direct bound $\Omega_{GW} h^{2} < 6\times 10^{25}$ in a narrow bandwidth around $100~\rm{MHz}$.  We should note however that the peaked GW spectrum predicted in this paper coming from inflaton fragmentation can easily occur at lower frequencies for low-scale models of inflation and reheating.  The only requirements for inflaton fragmentation are that the inflaton be complex (carrying a conserved $U(1)$ charge) and that the potential around the era of reheating be less steep than $\phi^{2}$.  Here we focus on chaotic inflation and leave a detailed survey of other models for future work.  

We note one further possible signature of high frequency GWs obtained from cosmic microwave background (CMB) measurements.  If the strength of primordial magnetic fields is sufficiently large, it may be possible to obtain indirect bounds on such GWs from CMB data~\cite{Pshirkov:2009sf}.  The observation of such GWs would provide an unprecedented view of the earliest moments of the Universe's history and energies well above those likely to be probed at particle colliders.

We begin in Section II with a general discussion of condensate fragmentation and $Q$-ball formation.  Although this discussion is general, we specialize to the case of running-mass chaotic inflation for the remainder of the paper for ease of calculation.  In Section III we analyze the rate of $Q$-ball decay which is an essential parameter governing the process of reheating. In Section IV we derive properties of the GW produced from the fragmentation. These properties include the peak amplitude and frequency as well as a schematic form of the complete GW spectrum. 
\section{Instabilities in the condensate}
In order  to illustrate this, let us consider a simple toy model of running mass inflaton with a negative Logarithmic correction, see for discussion~\cite{Mazumdar:2010sa}.  This may happen if the inflaton decay occurs through Yukawa couplings to fermions, i.e. $\mathcal{L}_{int}\sim h\phi\bar\psi\psi$  (where $\psi$ is the fermion, and $h$ is the Yukawa coupling)~\footnote{Inflaton coupling to another scalar will not lead to a complete decay of the inflaton. However, such coupling can trigger a phase of preheating via parametric resonance~\cite{GarciaBellido:2008rq} or a tachyonic instability~\cite{Mazumdar:2008up} which may produce additional GWs.  Here we will consider a minimal scenario for the inflaton to decay to fermions and study the consequences.  Further note that bounds on the Yukawa coupling to light fermions in running mass inflaton models are stringent~\cite{NeferSenoguz:2008nn}, but weak for heavy fermions. The couplings of the inflaton to these fermions need not directly reheat the Standard Model (SM) degrees of freedom as in the case of more realistic models of inflation, see~\cite{Allahverdi:2006we,Allahverdi:2006iq,Allahverdi:2006cx}. One might still require these fermions to eventually decay into the SM degrees of freedom which itself is a challenging problem.}. Otherwise the inflaton can have  gauge couplings in a supersymmetric theory, where the gaugino loops dominate to generate a one-loop  correction to the inflaton potential:
%
\be V(\phi) = m^{2} |\phi|^{2} \left(1 + K \log{ \frac{|\phi|^{2}}{M_{P}^{2}}}\right)+\cdots  \label{runmass} \ee
where $K$ can be positive or negative. Since in this paper our aim is to set a limit on the largest amplitude and frequency GW which can be produced after inflation, we would mainly concentrate on the first term during inflaton oscillations.
%
%

The potential Eq.~(\ref{runmass}) is consistent with the COBE normalization of the density perturbations if one takes 
%
$ {\delta \rho}/{\rho} \sim {m}/{M_{P}} \sim 10^{-5}$.
%
Implying that the inflaton mass must be about $m\sim 10^{13}~\rm{GeV}$ as noted above.  Note also that quadratic chatoic inflation predicts a spectral index $n_{s}=0.97$, in close agreement with the current WMAP data~\cite{Hinshaw:2008kr,NeferSenoguz:2008nn}. 

Typically, for matter couplings the value of $K$ is negative.  As a result when the inflaton oscillates coherently, it sees a shallower potential:
$V\sim m^2\phi^2 (\phi/M_P)^{2K}\propto \phi^{2+2K}$, which results in a negative pressure on average for an oscillating inflaton,
ie. $p\sim -(|K|/2)\rho$~\cite{Enqvist:1997si}. It is the presence of a negative pressure for certain sub-Huble modes that lead to their exponential growth and ultimate fragmentation of the condensate.

Although the process of fragmentation is non-perturbative~\cite{Kusenko:2009cv}, the largest instability mode can be tracked analytically in the linear perturbation analysis of the fluctuations of the inflaton condensate. Let us decompose the condensate into a spatially homogeneous part and a spacetime dependent fluctuation: $\phi \rightarrow \phi(t) + \delta \phi(x,t) $ and $\theta \rightarrow \theta(t) + \delta \theta(x,t)$. Then the classical equations of motion in a Fridemann-Robertson-Walker (FRW) universe 
%
%
%
for the inhomogeneous part are: 
\bea \ddot{\delta \theta} + 3H \dot{\delta \theta} - \frac{\nabla^2}{a^{2}}  \delta \theta + \frac{2 \dot{\phi}}{\phi} \dot{\delta \theta} + \frac{2 \dot{\theta}}{\theta} \dot{\delta \phi} - \frac{2 \dot{\phi} \dot{\theta}}{\phi^{2}} \delta \phi = 0, \label{theta}\\
\ddot{\delta \phi} + 3H \dot{\delta \phi} - \frac{\nabla^2}{a^{2}} \delta \phi - 2 \dot{\theta} \phi \dot{\delta \theta} + \delta \phi~ V''(\phi)  - \dot{\theta}^{2} \delta \phi =0,\label{modulus} \eea
%
where we have used $V'(\phi + \delta \phi) = V(\phi) + \delta \phi V''(\phi)$, which is accurate in the linear regime: $\delta \phi \ll \phi$. 
%
\be \delta \phi = \delta \phi_{0} e^{\alpha(t)+i kx}, ~~~\delta \theta = \delta \theta_{0} e^{\alpha(t) + ikx}. \label{ansatz}\ee
A solution that has real and positive $\alpha(t)$ is a signal that the condensate is unstable.  We can apply the ansatz (\ref{ansatz}) to equations (\ref{theta}) and (\ref{modulus}) to find a dispersion for $\alpha(t)$
\be \dot{\alpha}^{4} + \left(2 \frac{k^{2}}{a^{2}} + V''+ 3 \dot{\theta}^{2}\right) \dot{\alpha}^{2} + \left( \frac{k^{2}}{a^{2}} + V'' - \dot{\theta}^{2}\right) \frac{k^{2}}{a^{2}} = 0, \label{dispersion} \ee
 where we have assumed $\ddot{\alpha} \ll \dot{\alpha}^{2}$.  The positivity of $\dot{\alpha}(t)$ requires that the last term in Eq.~(\ref{dispersion}) be negative.  Thus we find the instability band to be 
\be 0 < \frac{k^{2}}{a^{2}} < \left( \dot{\theta}^{2} - V'' \right) \equiv \frac{k_{max}^{2}}{a^{2}}. \label{kbest} \ee
The best-amplified mode $k_{best}$ often lies close to the middle of the instability band~\cite{Kasuya:1999wu,Kasuya:2000wx, Kasuya:2000sc,Kasuya:2001hg}.  We can maximize $\dot{\alpha}(k)$ with respect to $k$ to find
\bea k_{best}^{2} = \frac{k_{max}^{2}}{16 \dot{\theta}^{2}} \left( V'' + 7 \dot{\theta}^{2}\right)
= k_{max}^{2} \left( \frac{1}{2} -\frac{k_{max}^{2}}{16 \dot{\theta}^{2}}, \right) 
\eea
where in the second line we used the definition of $k_{max}^2$ to eliminate $V''$. 

Using the fact that at an extremum $\dot{\alpha}(k)$ satisfies $\dot{\alpha}^{2}(k) = k_{max}^{2}/2 - k^{2}$, we find the rate of growth for the best-amplified mode as
\be \dot{\alpha}(k_{best})^{2} = \frac{k_{max}^{4}}{16 \dot{\theta}^{2}} \ee
In the limit that the Hubble expansion is negligible one can use the classical equation of motion for $\phi$ to show 
\be
\dot{\theta}^{2} \approx \frac{V'(\phi)}{\phi}\sim m^2[1+K+K\log(\phi^2/2M^2)] \ee
When this approximation is valid, one can write all the relevant expressions of the fragmentation $k_{best}$, $k_{max}$, and $\dot{\alpha}(k_{best})$ in terms of $V''$ and $V'$.  We list for completeness the parameters characterizing the fragmentation:
\bea 
&& k_{max}^{2} = \frac{V'}{\phi} - V'',~~~
k_{best}^{2} = k_{max}^{2} \left( \frac{1}{2} - \frac{k_{max}^{2}}{16} \frac{ \phi}{V'} \right), \nonumber \\
&&\dot{\alpha}(k_{best})^{2} = \frac{k_{max}^{4}}{16} \frac{\phi}{V'}\sim\frac{|K|^2m^2}{4}. \eea
These expressions emphasize the fact that when Hubble expansion is negligible and nonzero angular motion exists, the nature of the fragmentation is determined entirely by the potential and its derivatives.  Note that when the second derivative of the potential is small $V'' \ll V' /\phi$, $k_{max}^{2} \approx V' / \phi$. 

Let us note that the era of $Q$-ball formation is bounded by causality.  The best-amplified mode as given by Eq.~(11) for the running-mass potential is $k_{best}^{2} \approx m^{2} |K| \left( 1- |K|/4\right)$.  For this mode to be within the causally connected horizon, the size of the fragmented region cannot exceed that of the Hubble length at the time of fragmentation:
\be r^{-1}_{\ast}  \leq \pi^{-1/2}m |K|^{1/2} \sim 2 \times 10^{12}~\rm{GeV}\, \ee
for $m\sim 10^{13}$~GeV and $|K|\sim 0.1$. The actual time scale for the formation of lumps depends on the growth rate of 
forming Q-balls, which is roughly given by: 
\be t^{-1}_{f} \sim \gamma |K| m\,, \ee
where $\gamma\sim 0.1-0.001$ estimated from the numerical simulations~\cite{Enqvist:2002si,Kusenko:2009cv}. The estimated charge in each inflatonic $Q$-ball is roughly given by~\cite{Enqvist:2002si}
\bea Q = \frac{4\pi R^{3}}{3} n_{Q} &\approx& 4 \gamma^{2} |K|^{2} r_{\ast}^{3} m M_{P}^{2} \nonumber\\
&\approx& 7 \times 10^{6}.
\eea
with $\gamma \approx 10^{-2}$.  Since the mass of gravity-mediated-like $Q$-balls scales as $M(Q) \approx mQ$, these are generically unstable. As we will see in the next section, there are a number of processes that contribute to total rate of $Q$-ball decay.

\section{Decay of Q-balls}

The question of the rate of decay of these $Q$-ball is crucial to their impact on the GW signal. Moreover if the $Q$-balls do not decay sufficiently fast then they can come to dominate the energy density of the Universe.  The effect of a period of $Q$-ball domination on GW signals has been studied in the Affleck-Dine case by~\cite{Chiba:2009zu}.  Here however we study effects that preclude such a period.

Although we assume that the renormalizable scalar potential is $U(1)$ invariant,  there will in general be non-renormalizable operators indued by gravitational correction which violate this $U(1)$ symmetry. Of course there may be additional sources at low-energy physics with $U(1)$ violation, conservatively we can simply take that the scale of such $U(1)$ violation is $M \sim M_{P}$. Thus we include terms, such as 
\be V_{NR} (\phi) = \lambda \frac{\left(\phi^{n}+\phi^{*n}\right)}{M_P^{n-4}},\ee
Such terms lead to departure from purely radial motion of the complex field $\Phi = \phi e^{i \theta}$ in the internal $U(1)$ space, giving the condensate a nonzero $U(1)$ charge since $n_{Q} = \dot{\theta} |\phi|^{2}$~\cite{Allahverdi:2008pf}. The $U(1)$ violating interaction would break the condensate into smaller $Q$-balls~\cite{Kusenko:2009cv}. This process can proceed
perturbatively~\cite{Kusenko:2009cv} as well as non-perturbatively~\cite{Kawasaki:2005xc}. The lowest order operator $n=5$ would yield a decay rate, given by: $\Gamma_d \sim(1/Q)(dQ/dt)\sim \sigma n_{\phi}\sim \gamma^2|K|^2 (\phi_0/M_P)^2m$, where $\sigma \sim (1/M_P)^2$ and $n_{\phi}\sim \gamma^2|K|^2m\phi^2$. For the largest
amplitude of the oscillations, $\phi_0\leq M_P$, the fastest decay rate of the Q-balls is given by: $\Gamma_d \leq \gamma^2|K|^2m$. This time scale is fast but cannot prevent the initial inflaton fragmentation, which leads to the formation of $Q$-balls, i.e. $t_{f}^{-1}\gg \Gamma_d$. 

Furthermore, for a rotating inflaton condensate, there also exits the traditional perturbative decay channel~\cite{Allahverdi:2008pf,Allahverdi:2006xh} with a relatively slower rate $\Gamma\sim h^2m/8 \pi$. All of these would contribute to: lumps of inflatonic $Q$-balls and relativistic matter. The latter might not have come into full thermal equilibrium at initial stages, and would
have an initial momentum $\sim m$. This ambient plasma while interacting with $Q$-balls also would trigger rapid evaporation and diffusion of the $Q$-quanta from the surface area of the $Q$-ball. The diffusion rate is typically given by~\cite{Kasuya:2001hg}:
\be \Gamma_{diff} = - 4 \pi aT \ee
where $a \sim 4-6$ is a diffusion coefficient~\cite{Banerjee:2000mb}. The diffusion rate turns out to be the fastest way to get rid of the 
$Q$-balls  from the ambient plasma. For temperatures below the initial momentum of the relativistic fermions in the ambient plasma, i.e. $T<m \sim 10^{13}$~GeV and for $Q \lesssim 3 \times 10^{9}$ the diffusion rate leads to a complete evaporation of the $Q$-balls within one oscillation period$\sim m$. The decay of the $Q$-balls leads to the domination of relativistic fermions with an initial momentum of the order of the inflaton mass, which could be regarded as the instantaneous temperature of the relativistic plasma, $T_{inst}\sim m$.

\section{The spectrum of GW}

The time of maximal GW production and $Q$-ball formation both occur when the fluctuations go non-linear $\delta \phi /\phi \sim 1$.  Note that $Q$-balls form when $H_{f} \sim |K|^{1/2} m$  and oscillations begin around $H_{i} \sim m$.  We can therefore estimate the decay in the amplitude of the oscillations as $\phi_{0} \sim \phi_{i} (H_{f}/H_{i})$, where $\phi_{i} \sim M_{P}$, denotes the inflaton field value at the end of chaotic inflation.  This yields $\phi_{0} \sim |K| M_{P}$, which since $\delta \phi \sim \phi_{0}$, can be used to estimate the GW energy density at the time of fragmentation as
\be
\rho_{GW} = \frac{ |K|^{2} m^{2} \phi^{4}}{M_{P}^{2}}, \ee
such that
\bea \Omega_{GW}(t_{f}) =  \frac{\rho_{GW}}{m^{2} \phi_{0}^{2}} 
\approx 10^{-4} \left(\frac{|K|}{0.1}\right)^{4}.  \eea
By the present era this energy density has been substantially redshifted
\bea \Omega_{GW}(t_{0}) &=& \frac{\rho_{GW}(t_{0})}{\rho_{c}(t_{0})} =  \Omega_{GW}(t_{f}) \left(\frac{a_{f}}{a_{0}}\right)^{4} \left(\frac{H_{f}}{H_{0}}\right)^{2} \nonumber \\ 
&\approx& 10^{-5} h^{-2} \left(\frac{300}{g_{s,*}}\right)^{1/3}  \Omega_{GW}(t_{f}). 
\eea
Thus the present-day energy density fraction of GW from the fragmentation of the inflatonic condensate can be as large as $\Omega_{GW}h^{2}(t_{0}) \approx 10^{-9}$. 

The peak frequency of the gravitational wave spectrum will be largest when the condensate lumps are unstable and decay immediately on cosmological time scales.  In this case the peak frequency at the present epoch is related to the frequency at production time via the assumption of adiabatic evolution, $a(T) \propto g^{-1/3} T^{-1}$.  Numerical and analytic calculations~\cite{Kusenko:1997si,Kasuya:1999wu,Kasuya:2000wx, Kasuya:2000sc,Kasuya:2001hg,Enqvist:2002si} of the fragmentation process indicate that frequency of the fastest-growing mode is $f_{*} =  k_{best}/ 2\pi$ where $f_*^{-1}$ is the time of fragmentation~\cite{ Kusenko:2009cv}.  Then the frequency $f_{0} = f_{*} \left(a_{*}/a_{0}\right)$ is  
\bea f_{0} &=& f_{*} \left(\frac{a_{*}}{a_{inst}}\right) \left(\frac{g_{s,0}}{g_{s,R}}\right)^{1/3} \left(\frac{T_{0}}{T_{inst}}\right) \nonumber \\
&\approx &10^{10}\left(\frac{100}{g_{s,R}}\right)^{1/6}\left(\frac{T_{inst}}{10^{12}~{\rm GeV}}\right)~{\rm Hz},
 \eea
assuming $a_{*} \approx a_{inst}$, where all parameters with subscripts ``inst'' and ``0'' are evaluated at the epoch when the universe became radiation dominated with an instantaneous temperature, $T_{inst}\sim m$, and the present epoch respectively. 
This is by far the highest frequency GW 
which can be generated by any mechanism without affecting the scalar perturbations of the cosmic microave background radiation.



\begin{figure}
\centering
\epsfig{file=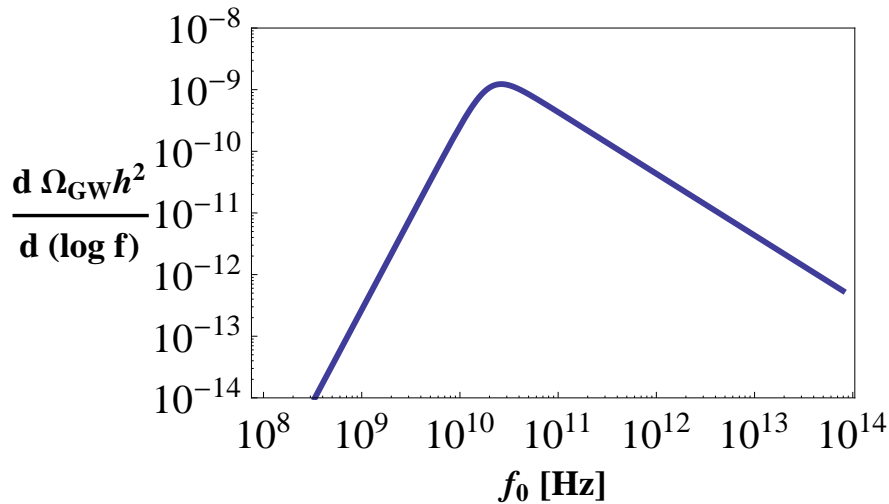,width=0.75\linewidth,clip=}
\caption{Here we plot the GW spectrum from inflaton fragmentation under the assumption that the fragmentation is an incoherent process, i.e. the anisotropic stresses are uncorrelated at different times.}
\label{spectrum}
\end{figure}
 The full spectrum can be estimated using existing methods for GW from first order phase transitions, see for instance~\cite{Caprini:2009yp,Chialva:2010jt}.  In this formalism one begins with the standard linearization of the Einstein equation 
 \be 
 \Box h_{ij} = \frac{32}{3} \pi G a^{2} \rho \Pi_{ij} \ee
 where $\rho$ is the energy density in the fragmenting condensate and $\Pi_{ij}$ is the anisotropic stress-energy tensor.  All the relevant information of the source is encoded in the power spectrum $P_{s}(k,t,t')$ which can be obtained from the anisotropic tensor via the 2-point correlation function
 \be \langle \Pi_{ij}(\mathbf{k},t) \Pi_{ij}(\mathbf{k'},t') \rangle = \left(2 \pi\right)^{3} \delta(\mathbf{k} - \mathbf{k'}) P_{s}(k,t,t'). \ee
 Using the definition of gravitational wave energy density $\rho_{GW}(\mathbf{x}) = \langle \dot{h}_{ij}(\mathbf{x}) \dot{h}^{*}_{ij}(\mathbf{x}) \rangle/( 8 \pi G a^{2})$ it can be shown that the present-day spectrum is given by
 \be \frac{d \Omega_{GW}}{d \log k}  = \frac{4 \Omega_{rad,0}}{3 \pi^{2}} \left(\frac{\Omega_{S}}{\Omega_{rad}}\right)^{2} H_{*}^{2} k^{3}\textsl{Re}[P_{s} (k,k,k,)], \label{spec} \ee
where we have assumed that the GW production occurs during radiation domination. Here we make the simplifying assumption that the power spectrum is separable at equal times $P_{s}(k,t,t) = |F(k)|^{2} |g(t)|^{2}$.  The time dependence of the fragmentation can be taken to turn on abruptly and last for a duration $\Gamma^{-1}$ such that $g(t)$ is unity during fragmentation and zero otherwise.  The spatial dependence $|F(k)|^{2}$ is taken to have the simple form
\be |F(k)|^{2} = \frac{R^{3}}{1+ \left(kR\right)^{4}} \ee
where $R = 2\pi k_{best}^{-1}$ is the characteristic scale of the fragmentation and the rate of fragmentation is approximately $\Gamma = \gamma^{-1} H_{*}$.  Moreover the fragmentation is a temporally random process. Thus in our case the GW source is totally incoherent in the sense that the anisotropic stress at different times is uncorrelated. Eq.~(\ref{spec}) is plotted in Fig. 1 under the assumption the fragmentation is temporally abrupt, short, and incoherent.

Before we conclude there are couple of points to note. The abundance of GW in the aftermath of inflation is the largest amongst any of the other competing scenarios. Scalar preheating can also create large amplitude GW with an abundance $\Omega_{GW}h^2\sim 10^{-10}$, but with a peak frequency $f\sim 10^{7}$~Hz~\cite{GarciaBellido:2008rq}, as compared to the fragmentation of the inflaton with $\Omega_{GW} h^{2} \sim 10^{-9}$ at peak frequencies around $10^{10}~\rm{Hz}$. In order to exceed the above limit one would require the universe to be filled with large energy densities and larger Hubble expansion rate. 

Note that in order to produce the large scale structures and the amplitude of the seed perturbations for the temperature anisotropy of order $10^{-5}$, the inflaton energy density can not be arbitrarily large, as it is bounded by the null observations of the stochastic GW which are inevitable consequence of inflation. However if future observations point towards peak frequencies in GWs exceeding $10^{10}$~Hz, then this would suggest a sub-Hubble origin for GWs at higher Hubble expansion rate and at higher energies then currently envisaged scale of inflation which is bounded by $10^{64}~({\rm GeV})^4$. This would perhaps be the cleanest way to falsify inflation as a source for seeding perturbations for the structure formation.

\section{Conclusions}
We have shown a simple and generic process that can produce extreme gravitational radiation with a sharply peaked spectrum, in contrast with what one ordinarily expects to produce during inflation.  The process of reheating in the early Universe is crucial for the production of entropy and the observed degrees of freedom.  Reheating is however a very difficult era to constrain by observation.  As we have shown above the if reheating involved an era of inflaton fragmentation, then it can have a critical impact on the prospects for GW detection. In particular, we have shown that a very early stage of inflaton fragmentation can produce extreme GWs with appreciable abundance in the multi-Gigahertz frequency range.  Such large frequencies make these GWs easily distinguishable from models of Affleck-Dine fragmentation, and the properties of the spectrum shown above make it distinct from what one would ordinarily predict from inflation. Although these GW are not amenable to observation via space-based interferometers it is conceivable that future tabletop experiments~\cite{Cruise:2000za,Cruise:2006zt} may be able to reach the required sensitivity in order to confirm or exclude an era of inflaton fragmentation in the early Universe.  Moreover, it may be possible to observe signatures from such GWs in the cosmic microwave background~\cite{Pshirkov:2009sf}.  Such an observation or refutation would give us profound insight into the nature of inflation and the era of reheating.

\acknowledgments

We are especially grateful to Alex Kusenko for very helpful discussions at the outset of this work. We would also like to thank Diego Chialva, Takeshi Chiba, Daniel Holz, Kohei Kamada, John McDonald, and Qaisar Shafi  for stimulating conversations and discussions.


\bibliography{qball}

\providecommand{\href}[2]{#2}\begingroup\raggedright\begin{thebibliography}{10}

\bibitem{Mazumdar:2010sa}
A.~Mazumdar and J.~Rocher, {\it {Particle physics models of inflation and
  curvaton scenarios}},  \href{http://xxx.lanl.gov/abs/1001.0993}{{\tt
  arXiv:1001.0993}}.

\bibitem{Hinshaw:2008kr}
{\bf WMAP} Collaboration, G.~Hinshaw {\em et.~al.}, {\it {Five-Year Wilkinson
  Microwave Anisotropy Probe (WMAP1 ) Observations:Data Processing, Sky Maps,
  \& Basic Results}},  {\em Astrophys. J. Suppl.} {\bf 180} (2009) 225--245,
  [\href{http://xxx.lanl.gov/abs/0803.0732}{{\tt arXiv:0803.0732}}].

\bibitem{Linde:1981mu}
A.~D. Linde, {\it {A New Inflationary Universe Scenario: A Possible Solution of
  the Horizon, Flatness, Homogeneity, Isotropy and Primordial Monopole
  Problems}},  {\em Phys. Lett.} {\bf B108} (1982) 389--393.

\bibitem{Linde:1983gd}
A.~D. Linde, {\it {Chaotic Inflation}},  {\em Phys. Lett.} {\bf B129} (1983)
  177--181.

\bibitem{Allahverdi:2010xz}
R.~Allahverdi, R.~Brandenberger, F.-Y. Cyr-Racine, and A.~Mazumdar, {\it
  {Reheating in Inflationary Cosmology: Theory and Applications}},
  \href{http://xxx.lanl.gov/abs/1001.2600}{{\tt arXiv:1001.2600}}.

\bibitem{Enqvist:2002si}
K.~Enqvist, S.~Kasuya, and A.~Mazumdar, {\it {Inflatonic solitons in running
  mass inflation}},  {\em Phys. Rev.} {\bf D66} (2002) 043505,
  [\href{http://xxx.lanl.gov/abs/hep-ph/0206272}{{\tt hep-ph/0206272}}].

\bibitem{Enqvist:2002rj}
K.~Enqvist, S.~Kasuya, and A.~Mazumdar, {\it {Reheating as a surface effect}},
  {\em Phys. Rev. Lett.} {\bf 89} (2002) 091301,
  [\href{http://xxx.lanl.gov/abs/hep-ph/0204270}{{\tt hep-ph/0204270}}].

\bibitem{Coleman:1985ki}
S.~R. Coleman, {\it {Q Balls}},  {\em Nucl. Phys.} {\bf B262} (1985) 263.

\bibitem{Kusenko:2008zm}
A.~Kusenko and A.~Mazumdar, {\it {Gravitational waves from fragmentation of a
  primordial scalar condensate into Q-balls}},
  \href{http://xxx.lanl.gov/abs/0807.4554}{{\tt arXiv:0807.4554}}.

\bibitem{Kusenko:2009cv}
A.~Kusenko, A.~Mazumdar, and T.~Multamaki, {\it {Gravitational waves from the
  fragmentation of a supersymmetric condensate}},
  \href{http://xxx.lanl.gov/abs/0902.2197}{{\tt arXiv:0902.2197}}.

\bibitem{Chiba:2009zu}
T.~Chiba, K.~Kamada, and M.~Yamaguchi, {\it {Gravitational Waves from Q-ball
  Formation}},  {\em Phys. Rev.} {\bf D81} (2010) 083503,
  [\href{http://xxx.lanl.gov/abs/0912.3585}{{\tt arXiv:0912.3585}}].

\bibitem{Chiba:2010ff}
T.~Chiba, K.~Kamada, S.~Kasuya, and M.~Yamaguchi, {\it {Fate of thermal log
  type Q balls}},  \href{http://xxx.lanl.gov/abs/1007.4235}{{\tt
  arXiv:1007.4235}}.

\bibitem{Kusenko:1997si}
A.~Kusenko and M.~E. Shaposhnikov, {\it {Supersymmetric Q-balls as dark
  matter}},  {\em Phys. Lett.} {\bf B418} (1998) 46--54,
  [\href{http://xxx.lanl.gov/abs/hep-ph/9709492}{{\tt hep-ph/9709492}}].

\bibitem{Kusenko:1997zq}
A.~Kusenko, {\it {Solitons in the supersymmetric extensions of the standard
  model}},  {\em Phys. Lett.} {\bf B405} (1997) 108,
  [\href{http://xxx.lanl.gov/abs/hep-ph/9704273}{{\tt hep-ph/9704273}}].

\bibitem{Enqvist:1997si}
K.~Enqvist and J.~McDonald, {\it {Q-balls and baryogenesis in the MSSM}},  {\em
  Phys. Lett.} {\bf B425} (1998) 309--321,
  [\href{http://xxx.lanl.gov/abs/hep-ph/9711514}{{\tt hep-ph/9711514}}].

\bibitem{Enqvist:1998ds}
K.~Enqvist and J.~McDonald, {\it {D-term inflation and B-ball baryogenesis}},
  {\em Phys. Rev. Lett.} {\bf 81} (1998) 3071--3074,
  [\href{http://xxx.lanl.gov/abs/hep-ph/9806213}{{\tt hep-ph/9806213}}].

\bibitem{Dine:2003ax}
M.~Dine and A.~Kusenko, {\it {The origin of the matter-antimatter asymmetry}},
  {\em Rev. Mod. Phys.} {\bf 76} (2004) 1,
  [\href{http://xxx.lanl.gov/abs/hep-ph/0303065}{{\tt hep-ph/0303065}}].

\bibitem{Enqvist:2003gh}
K.~Enqvist and A.~Mazumdar, {\it {Cosmological consequences of MSSM flat
  directions}},  {\em Phys. Rept.} {\bf 380} (2003) 99--234,
  [\href{http://xxx.lanl.gov/abs/hep-ph/0209244}{{\tt hep-ph/0209244}}].

\bibitem{Kasuya:1999wu}
S.~Kasuya and M.~Kawasaki, {\it {Q-ball formation through Affleck-Dine
  mechanism}},  {\em Phys. Rev.} {\bf D61} (2000) 041301,
  [\href{http://xxx.lanl.gov/abs/hep-ph/9909509}{{\tt hep-ph/9909509}}].

\bibitem{Kasuya:2000wx}
S.~Kasuya and M.~Kawasaki, {\it {Q-ball formation in the gravity-mediated SUSY
  breaking scenario}},  {\em Phys. Rev.} {\bf D62} (2000) 023512,
  [\href{http://xxx.lanl.gov/abs/hep-ph/0002285}{{\tt hep-ph/0002285}}].

\bibitem{Kasuya:2000sc}
S.~Kasuya and M.~Kawasaki, {\it {A new type of stable Q balls in the
  gauge-mediated SUSY breaking}},  {\em Phys. Rev. Lett.} {\bf 85} (2000)
  2677--2680, [\href{http://xxx.lanl.gov/abs/hep-ph/0006128}{{\tt
  hep-ph/0006128}}].

\bibitem{Kasuya:2001hg}
S.~Kasuya and M.~Kawasaki, {\it {Q-ball formation: Obstacle to Affleck-Dine
  baryogenesis in the gauge-mediated SUSY breaking?}},  {\em Phys. Rev.} {\bf
  D64} (2001) 123515, [\href{http://xxx.lanl.gov/abs/hep-ph/0106119}{{\tt
  hep-ph/0106119}}].

\bibitem{Frieman:1988ut}
J.~A. Frieman, G.~B. Gelmini, M.~Gleiser, and E.~W. Kolb, {\it {Solitogenesis:
  Primordial Origin of Nontopological Solitons}},  {\em Phys. Rev. Lett.} {\bf
  60} (1988) 2101.

\bibitem{Frieman:1989bx}
J.~A. Frieman, A.~V. Olinto, M.~Gleiser, and C.~Alcock, {\it {Cosmic Evolution
  of Nontopological Solitons. 1}},  {\em Phys. Rev.} {\bf D40} (1989) 3241.

\bibitem{Griest:1989bq}
K.~Griest and E.~W. Kolb, {\it {Solitosynthesis: Cosmological Evolution of
  Nontopological Solitons}},  {\em Phys. Rev.} {\bf D40} (1989) 3231.

\bibitem{Griest:1989cb}
K.~Griest, E.~W. Kolb, and A.~Massarotti, {\it {Statistical Fluctuations as the
  Origin of Nontopological Solitons}},  {\em Phys. Rev.} {\bf D40} (1989) 3529.

\bibitem{Enqvist:1998xd}
K.~Enqvist and J.~McDonald, {\it {MSSM dark matter constraints and decaying
  B-balls}},  {\em Phys. Lett.} {\bf B440} (1998) 59--65,
  [\href{http://xxx.lanl.gov/abs/hep-ph/9807269}{{\tt hep-ph/9807269}}].

\bibitem{Kusenko:1997ad}
A.~Kusenko, {\it {Small Q balls}},  {\em Phys. Lett.} {\bf B404} (1997) 285,
  [\href{http://xxx.lanl.gov/abs/hep-th/9704073}{{\tt hep-th/9704073}}].

\bibitem{Kusenko:1997hj}
A.~Kusenko, {\it {Phase transitions precipitated by solitosynthesis}},  {\em
  Phys. Lett.} {\bf B406} (1997) 26--33,
  [\href{http://xxx.lanl.gov/abs/hep-ph/9705361}{{\tt hep-ph/9705361}}].

\bibitem{Kusenko:1997it}
A.~Kusenko, M.~E. Shaposhnikov, P.~G. Tinyakov, and I.~I. Tkachev, {\it {Star
  wreck}},  {\em Phys. Lett.} {\bf B423} (1998) 104--108,
  [\href{http://xxx.lanl.gov/abs/hep-ph/9801212}{{\tt hep-ph/9801212}}].

\bibitem{Kusenko:1997vp}
A.~Kusenko, V.~Kuzmin, M.~E. Shaposhnikov, and P.~G. Tinyakov, {\it
  {Experimental signatures of supersymmetric dark-matter Q- balls}},  {\em
  Phys. Rev. Lett.} {\bf 80} (1998) 3185--3188,
  [\href{http://xxx.lanl.gov/abs/hep-ph/9712212}{{\tt hep-ph/9712212}}].

\bibitem{Laine:1998rg}
M.~Laine and M.~E. Shaposhnikov, {\it {Thermodynamics of non-topological
  solitons}},  {\em Nucl. Phys.} {\bf B532} (1998) 376--404,
  [\href{http://xxx.lanl.gov/abs/hep-ph/9804237}{{\tt hep-ph/9804237}}].

\bibitem{Enqvist:1998en}
K.~Enqvist and J.~McDonald, {\it {B-ball baryogenesis and the baryon to dark
  matter ratio}},  {\em Nucl. Phys.} {\bf B538} (1999) 321--350,
  [\href{http://xxx.lanl.gov/abs/hep-ph/9803380}{{\tt hep-ph/9803380}}].

\bibitem{Enqvist:1998pf}
K.~Enqvist and J.~McDonald, {\it {Observable isocurvature fluctuations from the
  Affleck-Dine condensate}},  {\em Phys. Rev. Lett.} {\bf 83} (1999)
  2510--2513, [\href{http://xxx.lanl.gov/abs/hep-ph/9811412}{{\tt
  hep-ph/9811412}}].

\bibitem{Axenides:1999hs}
M.~Axenides, S.~Komineas, L.~Perivolaropoulos, and M.~Floratos, {\it {Dynamics
  of nontopological solitons: Q balls}},  {\em Phys. Rev.} {\bf D61} (2000)
  085006, [\href{http://xxx.lanl.gov/abs/hep-ph/9910388}{{\tt
  hep-ph/9910388}}].

\bibitem{Banerjee:2000mb}
R.~Banerjee and K.~Jedamzik, {\it {On B-ball dark matter and baryogenesis}},
  {\em Phys. Lett.} {\bf B484} (2000) 278--282,
  [\href{http://xxx.lanl.gov/abs/hep-ph/0005031}{{\tt hep-ph/0005031}}].

\bibitem{Battye:2000qj}
R.~Battye and P.~Sutcliffe, {\it {Q-ball dynamics}},  {\em Nucl. Phys.} {\bf
  B590} (2000) 329--363, [\href{http://xxx.lanl.gov/abs/hep-th/0003252}{{\tt
  hep-th/0003252}}].

\bibitem{Allahverdi:2002vy}
R.~Allahverdi, A.~Mazumdar, and A.~Ozpineci, {\it {Q-ball formation in the wake
  of Hubble-induced radiative corrections}},  {\em Phys. Rev.} {\bf D65} (2002)
  125003, [\href{http://xxx.lanl.gov/abs/hep-ph/0203062}{{\tt
  hep-ph/0203062}}].

\bibitem{Kusenko:2004yw}
A.~Kusenko, L.~Loveridge, and M.~Shaposhnikov, {\it {Supersymmetric dark matter
  Q-balls and their interactions in matter}},  {\em Phys. Rev.} {\bf D72}
  (2005) 025015, [\href{http://xxx.lanl.gov/abs/hep-ph/0405044}{{\tt
  hep-ph/0405044}}].

\bibitem{Kusenko:2005du}
A.~Kusenko, L.~C. Loveridge, and M.~Shaposhnikov, {\it {Astrophysical bounds on
  supersymmetric dark-matter Q- balls}},  {\em JCAP} {\bf 0508} (2005) 011,
  [\href{http://xxx.lanl.gov/abs/astro-ph/0507225}{{\tt astro-ph/0507225}}].

\bibitem{Berkooz:2005rn}
M.~Berkooz, D.~J.~H. Chung, and T.~Volansky, {\it {High density preheating
  effects on Q-ball decays and MSSM inflation}},  {\em Phys. Rev. Lett.} {\bf
  96} (2006) 031303, [\href{http://xxx.lanl.gov/abs/hep-ph/0510186}{{\tt
  hep-ph/0510186}}].

\bibitem{Berkooz:2005sf}
M.~Berkooz, D.~J.~H. Chung, and T.~Volansky, {\it {Constraining modular
  inflation in the MSSM from giant Q- ball formation}},  {\em Phys. Rev.} {\bf
  D73} (2006) 063526, [\href{http://xxx.lanl.gov/abs/hep-ph/0507218}{{\tt
  hep-ph/0507218}}].

\bibitem{Johnson:2008se}
M.~C. Johnson and M.~Kamionkowski, {\it {Dynamical and Gravitational
  Instability of Oscillating- Field Dark Energy and Dark Matter}},
  \href{http://xxx.lanl.gov/abs/0805.1748}{{\tt arXiv:0805.1748}}.

\bibitem{Kasuya:2008xp}
S.~Kasuya, M.~Kawasaki, and F.~Takahashi, {\it {Isocurvature fluctuations in
  Affleck-Dine mechanism and constraints on inflation models}},
  \href{http://xxx.lanl.gov/abs/0805.4245}{{\tt arXiv:0805.4245}}.

\bibitem{Sakai:2007ft}
N.~Sakai and M.~Sasaki, {\it {Stability of Q-balls and Catastrophe}},  {\em
  Prog. Theor. Phys.} {\bf 119} (2008) 929--937,
  [\href{http://xxx.lanl.gov/abs/0712.1450}{{\tt arXiv:0712.1450}}].

\bibitem{Campanelli:2007um}
L.~Campanelli and M.~Ruggieri, {\it {Supersymmetric Q-balls: A Numerical
  Study}},  {\em Phys. Rev.} {\bf D77} (2008) 043504,
  [\href{http://xxx.lanl.gov/abs/0712.3669}{{\tt arXiv:0712.3669}}].

\bibitem{Kawasaki:2005xc}
M.~Kawasaki, K.~Konya, and F.~Takahashi, {\it {Q-ball instability due to U(1)
  breaking}},  {\em Phys. Lett.} {\bf B619} (2005) 233,
  [\href{http://xxx.lanl.gov/abs/hep-ph/0504105}{{\tt hep-ph/0504105}}].

\bibitem{Kasuya:2007cy}
S.~Kasuya and F.~Takahashi, {\it {Entropy production by Q-ball decay for
  diluting long-lived charged particles}},  {\em JCAP} {\bf 0711} (2007) 019,
  [\href{http://xxx.lanl.gov/abs/0709.2634}{{\tt arXiv:0709.2634}}].

\bibitem{Shoemaker:2008gs}
I.~M. Shoemaker and A.~Kusenko, {\it {The ground states of baryoleptonic
  Q-balls in supersymmetric models}},  {\em Phys. Rev.} {\bf D78} (2008)
  075014, [\href{http://xxx.lanl.gov/abs/0809.1666}{{\tt arXiv:0809.1666}}].

\bibitem{Campanelli:2009su}
L.~Campanelli and M.~Ruggieri, {\it {Spinning Supersymmetric Q-balls}},  {\em
  Phys. Rev.} {\bf D80} (2009) 036006,
  [\href{http://xxx.lanl.gov/abs/0904.4802}{{\tt arXiv:0904.4802}}].

\bibitem{Shoemaker:2009kg}
I.~M. Shoemaker and A.~Kusenko, {\it {Gravitino dark matter from Q-ball
  decays}},  {\em Phys. Rev.} {\bf D80} (2009) 075021,
  [\href{http://xxx.lanl.gov/abs/0909.3334}{{\tt arXiv:0909.3334}}].

\bibitem{Kusenko:2009iz}
A.~Kusenko and I.~M. Shoemaker, {\it {Neutrinos from the terrestrial passage of
  supersymmetric dark matter Q-balls}},  {\em Phys. Rev.} {\bf D80} (2009)
  027701, [\href{http://xxx.lanl.gov/abs/0905.3929}{{\tt arXiv:0905.3929}}].

\bibitem{Cruise:2000za}
A.~M. Cruise, {\it {An electromagnetic detector for very-high-frequency
  gravitational waves}},  {\em Class. Quant. Grav.} {\bf 17} (2000) 2525--2530.

\bibitem{Cruise:2006zt}
A.~M. Cruise and R.~M.~J. Ingley, {\it {A prototype gravitational wave detector
  for 100-MHz}},  {\em Class. Quant. Grav.} {\bf 23} (2006) 6185--6193.

\bibitem{Nishizawa:2007tn}
A.~Nishizawa {\em et.~al.}, {\it {Laser-interferometric Detectors for
  Gravitational Wave Background at 100 MHz : Detector Design and Sensitivity}},
   {\em Phys. Rev.} {\bf D77} (2008) 022002,
  [\href{http://xxx.lanl.gov/abs/0710.1944}{{\tt arXiv:0710.1944}}].

\bibitem{Nishizawa:2008se}
A.~Nishizawa {\em et.~al.}, {\it {Optimal Location of Two Laser-interferometric
  Detectors for Gravitational Wave Backgrounds at 100 MHz}},  {\em Class.
  Quant. Grav.} {\bf 25} (2008) 225011,
  [\href{http://xxx.lanl.gov/abs/0801.4149}{{\tt arXiv:0801.4149}}].

\bibitem{Akutsu:2008qv}
T.~Akutsu {\em et.~al.}, {\it {Search for a stochastic background of 100-MHz
  gravitational waves with laser interferometers}},  {\em Phys. Rev. Lett.}
  {\bf 101} (2008) 101101, [\href{http://xxx.lanl.gov/abs/0803.4094}{{\tt
  arXiv:0803.4094}}].

\bibitem{Pshirkov:2009sf}
M.~S. Pshirkov and D.~Baskaran, {\it {Limits on High-Frequency Gravitational
  Wave Background from its interplay with Large Scale Magnetic Fields}},  {\em
  Phys. Rev.} {\bf D80} (2009) 042002,
  [\href{http://xxx.lanl.gov/abs/0903.4160}{{\tt arXiv:0903.4160}}].

\bibitem{GarciaBellido:2008rq}
J.~Garcia-Bellido and D.~G. Figueroa, {\it {A new gravitational wave background
  from the Big Bang}},  \href{http://xxx.lanl.gov/abs/0801.4109}{{\tt
  arXiv:0801.4109}}.

\bibitem{Mazumdar:2008up}
A.~Mazumdar and H.~Stoica, {\it {Exciting gauge field and gravitons in a
  brane-anti-brane annihilation}},  {\em Phys. Rev. Lett.} {\bf 102} (2009)
  091601, [\href{http://xxx.lanl.gov/abs/0807.2570}{{\tt arXiv:0807.2570}}].

\bibitem{NeferSenoguz:2008nn}
V.~N. Senoguz and Q.~Shafi, {\it {Chaotic inflation, radiative corrections and
  precision cosmology}},  {\em Phys. Lett.} {\bf B668} (2008) 6,
  [\href{http://xxx.lanl.gov/abs/0806.2798}{{\tt arXiv:0806.2798}}].

\bibitem{Allahverdi:2006we}
R.~Allahverdi, K.~Enqvist, J.~Garcia-Bellido, A.~Jokinen, and A.~Mazumdar, {\it
  {MSSM flat direction inflation: slow roll, stability, fine tunning and
  reheating}},  {\em JCAP} {\bf 0706} (2007) 019,
  [\href{http://xxx.lanl.gov/abs/hep-ph/0610134}{{\tt hep-ph/0610134}}].

\bibitem{Allahverdi:2006iq}
R.~Allahverdi, K.~Enqvist, J.~Garcia-Bellido, and A.~Mazumdar, {\it {Gauge
  invariant MSSM inflaton}},  {\em Phys. Rev. Lett.} {\bf 97} (2006) 191304,
  [\href{http://xxx.lanl.gov/abs/hep-ph/0605035}{{\tt hep-ph/0605035}}].

\bibitem{Allahverdi:2006cx}
R.~Allahverdi, A.~Kusenko, and A.~Mazumdar, {\it {A-term inflation and the
  smallness of the neutrino masses}},  {\em JCAP} {\bf 0707} (2007) 018,
  [\href{http://xxx.lanl.gov/abs/hep-ph/0608138}{{\tt hep-ph/0608138}}].

\bibitem{Allahverdi:2008pf}
R.~Allahverdi and A.~Mazumdar, {\it {Affleck-Dine condensate, late
  thermalization and the gravitino problem}},  {\em Phys. Rev.} {\bf D78}
  (2008) 043511, [\href{http://xxx.lanl.gov/abs/0802.4430}{{\tt
  arXiv:0802.4430}}].

\bibitem{Allahverdi:2006xh}
R.~Allahverdi and A.~Mazumdar, {\it {Longevity of supersymmetric flat
  directions}},  {\em JCAP} {\bf 0708} (2007) 023,
  [\href{http://xxx.lanl.gov/abs/hep-ph/0608296}{{\tt hep-ph/0608296}}].

\bibitem{Caprini:2009yp}
C.~Caprini, R.~Durrer, and G.~Servant, {\it {The stochastic gravitational wave
  background from turbulence and magnetic fields generated by a first-order
  phase transition}},  {\em JCAP} {\bf 0912} (2009) 024,
  [\href{http://xxx.lanl.gov/abs/0909.0622}{{\tt arXiv:0909.0622}}].

\bibitem{Chialva:2010jt}
D.~Chialva, {\it {Gravitational waves from first order phase transitions during
  inflation}},  \href{http://xxx.lanl.gov/abs/1004.2051}{{\tt
  arXiv:1004.2051}}.

\end{thebibliography}\endgroup
\bibliographystyle{JHEP}

\end{document}